# The Fate of Evolutionary Archaeology: Survival or Extinction?

Liane Gabora

Address for correspondence:

Liane Gabora
University of British Columbia
Okanagan campus, 3333 University Way
Kelowna BC, V1V 1V7, CANADA
Email: liane.gabora@ubc.ca
Phone: (250) 807-9849
Fax: (250) 470-6001

**Abstract**
It is important to be clear as to whether a theory such as evolutionary archaeology pertains to biological evolution, in which acquired change is obliterated at the end of each generation, or cultural change, in which acquired change is retained. In evolutionary archaeology, (1) the population is said to consist of artifacts, yet (2) artifacts are said to be phenotypic. Neither (1) nor (2) is necessarily problematic in and of itself, but the two are inconsistent, as the first pertains to cultural change whereas the second to the biological evolution of humans. A first step to avoiding this problem is to recognize that there is a need for a theory of change specific to human culture. Referring to ongoing work using a related approach to cultural change, it is suggested that the inconsistencies in evolutionary archaeology, though problematic, are not insurmountable.

Artifacts undeniably exhibit identifiable patterns of change over time, or 'descent with' modification, as do biological organisms. The goal behind an evolutionary approach to archaeology seems reasonable and straightforward: to develop an overarching framework for analyzing how artifacts are related to one another, borrowing concepts developed for this purpose in biology such as lineage, heritability, and drift (changes in the relative frequencies of different alleles—forms of a gene—as a statistical byproduct of randomly sampling from a finite population). This kind of framework would be indeed be of value, and those who have striven to go beyond their primary area of expertise to build such a framework are to be commended, for it is easy to underestimate the difficulty of such an endeavor. However, much of this literature is riddled with serious conceptual errors. Some have been debated at length (*e.g*. Bamforth 2002; Boone and Smith 1998; Kehoe 2000; Loney 2000; Lyman and O'Brien 1997; Murray 2002; Neff 2000; O'Brien 1996a, b, 2005; O'Brien and Lyman 2004; Preucel 1999; Schiffer 1996; Shennan 2002; Spencer 1997; Wylie 2000). Others, however, have not been addressed, or require elaboration. The goal of this paper is to tackle these errors because, as I see it, they are exactly what stands in the way to a genuine theory of how artifacts evolve.

**If Artifacts are the Phenotype then Humans are the Population**

Let us jump right in by addressing a glaring problem that comes to light when evolutionary archaeologists are charged with not having a clear answer as to what the unit of replication is in the evolution of artifacts. Lyman and O'Brien answer that Darwin did not know what the unit of replication was when he came up with the theory of how organisms evolve through natural selection, so they needn't be concerned that they don't yet know what the unit of replication is with respect to the evolution of artifacts (1998:619). However, Darwin *did* know what the unit of *replication* is: the organism. It is the unit of *heredity*—the gene—that he did not know.

This lack of clarity as to what is the unit of replication leads quite naturally to a persistent error as to what is the population. O'Brien claims "in archaeology, not surprisingly, the population consists of artifacts" (2005:30). But this *is* surprising, for earlier in the same paper he wrote:

> The only defense that Brew could see for even thinking of using an artifact-classification system 'based upon phylogenetic theory is that the individual objects were made and used by man' (1946:55)—a point that, to Brew at least, was so obvious as to be trivial. As we shall see, however, that point is the keystone to applying Darwinism to the archaeological record. (2005:28)

This leads one to believe that O'Brien's position is that it is because humans evolve through natural selection that Darwinian theory is relevant to artifacts. But in the evolution of humans through natural selection, the relevant population consists of *humans*, not artifacts. Indeed this interpretation of his position is strengthened a few lines after the statement that the population is artifacts, where he writes:

> Evolutionary archaeology rests on the premise that objects in the archaeological record, because they were parts of past phenotypes, were shaped by the same evolutionary processes as were the somatic (bodily) features of their makers and users. (2005:30)

Here it is unmistakable that O'Brien is speaking of biological evolution operating on populations of humans rather than cultural evolution operating on populations of artifacts. Elsewhere this inconsistency appears even within a single sentence, for instance "In evolutionary archaeology, the population is artifacts, which are viewed as phenotypic features…" (Lyman and O'Brien, 1997: p. 616). To be consistent with the first part of the sentence, 'the population is artifacts', each artifact must be understood to possess a unique genotype, or something that plays the role of a genotype, which gets expressed phenotypically. However the second part claims that artifacts are 'viewed as phenotypic features'. How can something be both (1) an entity with a phenotypically expressed genotype, and (2) a mere phenotypic feature? One gets the impression that in the second part of the sentence Lyman and O'Brien mean that artifacts are phenotypic features of the human genotype. But then once again the relevant population is the set of interbreeding humans, not artifacts. It is important to see here that the problem is *not* that they consider artifacts part of the human phenotype. That is not at issue. The issue is that the framework they are constructing is *internally* inconsistent.

**Why it Matters that Artifacts Inherit Acquired Characteristics**

The confusion regarding the unit of replication and the population are all the more problematic because the mechanisms underlying the biological evolution of humans and the cultural evolution of artifacts are different, as evidenced by the fact that while biological organisms are protected from change accrued during a lifetime (for example if one cuts off the tail of a rat, its offspring still have tails of normal length) for artifacts this is not the case. Once someone made a cup with a handle, cups with handles were here to stay. Indeed in human culture, inheritance of acquired characteristics is not just the exception but the rule (though note that here 'inherited' merely means transmitted or 'passed on' without implying genetic mediation). Characteristics of artifacts actually change faster than the genomes of the individuals who produce them. The mathematical theory of natural selection developed to describe biological evolution[1] is only applicable when there is no inheritance of acquired characteristics (or at least it must be negligible compared to change due to differential replication of individuals with heritable variation competing for scarce resources), a condition clearly not met in the evolution of culture.

Since biology and culture are different (though intertwined) processes with different underlying mechanisms, it is important to be clear which of them one is theorizing about: the one where inheritance of acquired characteristics is prohibited (biology), or the one where it is not (culture). The EA literature is rampant with passages that reveal a lack of appreciation of this key point. Recall Brew's objection to applying Darwinism to artifacts merely because they are made by humans. What Brew understood (either rationally or intuitively), and what O'Brien missed, is that since artifacts can change much more rapidly than the genomes of the individuals who make them, change in artifacts cannot be explained by recourse to the fact that humans are evolving through natural selection. It is important to emphasize that no one disputes that humans evolve through natural selection; on this point, O'Brien's characterization of his critics is distorted. No one ever argued for instance that "intelligence and motivation exempt humans from the evolutionary process" (2005:30-31) or that "they are not subject to natural selection and drift" (2005:31) or that "adaptive plasticity shields humans from evolutionary processes" (2005:31) or that "humans can dodge natural selection by making choices" (2005:31). That would indeed be ridiculous. The argument is subtler, and can perhaps be understood through the following analogy. One can use knowledge about the wind to make predictions about the speed and position of an unmanned boat adrift at sea. But if the boat is manned and equipped with oars or a motor, wind may not be the major causal factor determining its speed and position. It is not that the boat is no longer affected by wind; it is that some other causal factor is playing an influential role.

[1] This was initiated by the pioneering efforts of Fisher (2930), Haldane (1932), and Wright (1931), and has since been extensively developed by other population geneticists.

**Special Laws for Humans?**

EA's refusal to accept this appears to stem from the unfounded presumption that it is incompatible with human behavior as being phenotypic. As O'Brien and Lyman (2004) put it:

> That artifacts are phenotypic is nonproblematic to most biologists, who routinely view such things as a bird's next, a beaver's dam, or a chimpanzee's twig tools as phenotypic traits… If the behaviors are phenotypic then the results of the behaviors are phenotypic as well. But many of us have a problem with viewing human behavior as phenotypic. We see ourselves as being quantitatively and/or qualitatively *so* different from the rest of the natural world that we warrant not only a whole new set of laws but also a different set of philosophical questions with which to examine ourselves. EA does not accept that argument. (p. 179)

Their use of the word 'laws' is misleading, as if a phenomenon has to attain the status of the 'law of gravity' to be worthy of attention. It is ironic though, in a sense, that they chose this word, because the subject of law *does* address an immense body of questions and issues pertaining solely to humans. The law does not prescribe rules of conduct for rats or rocks. And of course, law is not unique in this respect. At least as many disciplines deal exclusively with human affairs (political science, literature, sociology…) as with affairs not restricted to the human realm (physics, chemistry, geology…).

O'Brien and Lyman repeatedly claim that their approach is more scientific than alternatives, stressing "Hypotheses derived from theories must have testable implications (2004, p. 187)." In arguing that natural selection at the biological level provides all the explanatory power we need, we are led to the hypothesis that we make artifacts like popcorn makers and Powerpuff dolls because we are genetically predisposed to, and if this is born out I will have newfound admiration for this approach. However I suspect that, much as physics does not get us far toward an explanation of, say, predator-prey relationships, biology will not go far toward an explanation of the design of cultural artifacts.

The indisputable fact is that we *are* different from the rest of the natural world, both qualitatively and quantitatively. Nothing is gained by letting ideals (concerning the equality of all forms of life) obscure the evidence before our eyes that we (for better or worse) have abilities other species lack. Considerable effort has gone into attempting to explain human functioning and behavior in terms of theories that have proven successful for explaining the natural world. And these efforts have been successful to the extent that humans share characteristics with other aspects of the natural world; for example in understanding how the eye works or how disease spreads or even aspects of courtship behavior. But these efforts have not taken us very far when it comes to explaining characteristics by which we *differ* from other elements of the natural world. And what makes us most unique is exactly that aspect of human nature that is of relevance to archaeology: our propensity to not just generate novelty but to build on it cumulatively, adapting old ideas to new circumstances (the Ratchet effect). The upshot, as Renfrew (1982) pointed out some time ago, is that genetically inherited traits likely play but a minor role in the form of a human-made artifact.

This does not imply that artifacts do not affect biological fitness, or that biological constraints do not affect their design. It just means that even if you had a complete understanding of biology, and in particular of genes and how they are expressed in different environments, you would not have enough information to predict or even interpret change in the archaeological record. Moreover, archaeological data may still give evidence of phenomena observed in biology such as lineages and drift. But when such phenomena are observed they need not be *attributed* to natural selection. Indeed drift and other population phenomena have been observed in a computer model of cultural evolution in which neural network based agents (without anything playing the role of genomes) evolve gestures through a re-iterated non-Darwinian process of invention and imitation, no natural selection at all (Gabora 1995).

**Conclusions**

This paper is far from the first to point out flaws with EA. Attacks have come from multiple directions, and addressed multiple aspects of its claims. But that does not mean that EA is doomed to extinction. Perhaps evolutionary theory does have a role to play in a workable theory of artifact change, and the goal of constructing and analyzing artifact lineages using concepts from biology need not be forsaken, but it is time for proponents of EA to cut their losses and rethink their strategy.

A first step is to resolve conceptual inconsistencies (outlined above) concerning the unit of replication, the population, the inheritance of acquired characteristics, and so forth. A next step is to forgo that biological evolution can provide a complete explanation for artifact change. Moreover, it may be that the cultural process overlying it is quite different from how it is now conceived. In the scheme I put forth, the unit of replication in culture is the integrated web of memories, concepts, ideas, and attitudes that constitute an internal model of the world and how to conduct oneself in it: a worldview (Gabora 1998, 1999, 2000, 2004; Gabora and Aerts 2005a). Like the autocatalytic sets widely thought to be the earliest forms of life (Gabora 2006; Kauffman 1993), worldviews replicate through an emergent, autopoietic process, without a self-assembly code (like the genetic code), and thus acquired characteristics are inherited. Novelty is introduced not randomly but through strategic, creative processes (Gabora 2005). Artifacts mediate the process by which ones' worldview is transmitted, in a clumsy, piecemeal manner, to others. One cannot examine the state of a worldview, but much like footprints in the snow tell us much about the animal that left them, artifacts tell us much about the minds that created them.

So is the fate of EA survival or extinction? The question itself assumes there are only these two possibilities, which is consistent with the Darwinian perspective espoused by evolutionary archaeologists, in which only the fittest survive. However, in the alternative proposed above, where culture evolves through a process more akin to the evolution of pre-DNA life than post-DNA life—a process that in fact *violates* the assumptions that make natural selection applicable to its description (Gabora 2006; Gabora and Aerts 2005b)—survival and extinction are not the only two possible fates. Because uncoded replicators inherit acquired traits, they may either survive intact or go extinct, or do something in between these two extremes: *transform*. EA may have to do considerable transforming to stay afloat. But if the missing 'traits' are 'acquired', vestiges of the enterprise as currently practiced and advocated may linger in theories that float above it.

**References**

Bamforth, D. B. 2002. Evolution and metaphor in evolutionary archaeology. *American Antiquity*, 67: 435–52.

Boone, J. L., Smith, E. A. 1998. Is it evolution yet? A Critique of Evolutionary Archaeology. *Current Anthropology,* 39: S141–73.

Fisher, R. A. 1930. *The genetical theory of natural selection.* Clarendon: Oxford University Press.

Gabora, L. 1995. Meme and Variations: A computer model of cultural evolution. In *1993 Lectures in Complex Systems* (eds. Nadel, L., Stein, D.), Reading MA: Addison-Wesley, pp. 471–486.

Gabora, L. 1998. A tentative scenario for the origin of culture. *Psycoloquy,* ***9***(67).

Gabora, L. 2000. Conceptual closure: Weaving memories into an interconnected worldview. In *Closure: Emergent Organizations and their Dynamics* (eds. Van de Vijver, G., Chandler, J.). Annals of the New York Academy of Sciences 901, New York.

Gabora, L. 2004. Ideas are not replicators but minds are. *Biology & Philosophy,* 19(1): 127–143.

Gabora, L. 2005. Mind: What archaeology can tell us about the origins of human cognition. In *Handbook of Theories and Methods in Archaeology* (eds. A. Bentley and H. Maschner). Walnut Creek CA: Altamira Press.

Gabora, L. 2005. Creative thought as a non-Darwinian evolutionary process. *Journal of Creative Behavior*, 39(4): 65–87.

Gabora, L., in press. Self-other organization: Why early life did not evolve through natural selection. *Journal of Theoretical Biology*. (DOI: 10.1016/j.jtbi.2005.12.007)

Gabora, L. & Aerts, D. 2005. Distilling the essence of an evolutionary process, and implications for a formal description of culture. In *Proceedings of Center for Human Evolution Workshop #5: Cultural Evolution, May 2000* (ed. W. Kistler). Foundation for the Future, Seattle WA.

Gabora, L. & Aerts, D. 2005. Evolution as context-driven actualization of potential: Toward an interdisciplinary theory of change of state. *Interdisciplinary Science Reviews*, ***30***(1), 69-88.

Haldane, J. B. S. 1932. *The Causes of Evolution.* New York: Longman.

Kauffman, S. A. 1993. *Origins of Order*. New York: Oxford University Press.

Kehoe, A. B. 2000. Evolutionary archaeology challenges the future of archaeology: Response to O'Brien and Lyman. *Review of Archaeology*, 21: 33–38.

Loney, H. L. 2000. Society and technological control: A critical review of models of technological change in ceramic studies. *American Antiquity*, 65: 646–668.

Lyman, R. L. and O'Brien, M. J. 1997. Goals of evolutionary archaeology: history and explanation. *Current Anthropology,* 39(5): 615–652.

Murray, T. 2002. Evaluating evolutionary archaeology. World Archaeology, 34(1): 47–59.

Neff, H. 2000. On evolutionary ecology and evolutionary archaeology: Some common ground? *Current Anthropology*, 41: 427–429.

O'Brien, M. J. 1996a. The historical development of an evolutionary archaeology. In *Darwinian Archaeologies* (ed. H. D. G. Maschner). New York: Plenum Press, pp. 17–32.

O'Brien, M. J. 1996b. Evolutionary archaeology: An introduction. In *Evolutionary Archaeology: Theory and Application* (ed. M. J. O'Brien). Salt Lake City: University of Utah Press, pp. 1–15.

O'Brien, M. J. 2005. Evolutionism and North American's archaeological record. *World Archaeology*, 37(1): 26-45.

O'Brien, M. J., Lyman, R. L. 2004. History and explanation in archaeology. *Anthropological Theory,* 4(2): 173–197.

Preucel, R. W. 1999. Review of "Evolutionary Archaeology: Theory and Application". *Journal of Field Archaeology*, 26: 93–99.

Renfrew, C. 1982. *Towards an Archaeology of Mind: An Inaugural Lecture*. Cambridge: Cambridge University Press.

Schiffer, M. B. 1996. Some relationships between behavioral and evolutionary archaeologies. *American Antiquity*, 61: 643–662.

Shennan, S. 2002. Archaeology evolving: History, adaptation, self-organization. *Antiquity*, 76: 253–256.

Spencer, C. S. 1997. Evolutionary approaches in archaeology. *Journal of Archaeological Research,* 5: 209–264.

Wright, S. 1931. Evolution in Mendelian populations. *Genetics*, 16, 97–159.

Wylie, W. 2000. Questions of Evidence, Legitimacy, and the (Dis)Unity of Science. *American Antiquity*, 65(2): 227-237.